\documentclass[twocolumn,showpacs,preprintnumbers,amsmath,amssymb]{revtex4}

\usepackage{graphicx}
\usepackage{dcolumn}
\usepackage{bm}

\begin{document}


\title{Derivation of evolutionary payoffs from observable behavior.}
\author{}
\author{Alexander Feigel\footnote{Electronic address: sasha@soreq.gov.il}, Avraham Englander and Assaf Engel}
\affiliation{%
Soreq NRC,\\
Yavne 81800, Israel}


\date{\today}

\begin{abstract}

Interpretation of animal behavior, especially as cooperative or selfish, is a challenge for evolutionary theory. Strategy of a competition should follow from corresponding Darwinian payoffs for the available behavioral options. The payoffs and decision making processes, however, are difficult to observe and quantify. Here we present a general method for the derivation of evolutionary payoffs from observable statistics of interactions. The method is applied to combat of male bowl and doily spiders, to predator inspection by sticklebacks and to territorial defense by lions, demonstrating animal behavior as a new type of game theoretical equilibrium. Games animals play may be derived unequivocally from their observable behavior, the reconstruction, however, can be subjected to fundamental limitations due to our inability to observe all information exchange mechanisms (communication).

\end{abstract}

\maketitle

Animals in the wild or laboratory environment are involved in coordinated, game-like interactions. It can be a fight for mating opportunities with basic individual choices to attack or retreat, or cooperation vs. selfishness dilemma such as participation in territory defence or predator inspection. The corresponding Darwinian rewards and costs are significant, exacting even the life of a player. Animal behavior in these contests, therefore, must be affected by natural selection, converging with time to some evolutionary optimum.

Modeling of animal behavior in the framework of evolutionary game theory\cite{Smith1982,Vincent2005,Nowak2004}, requires a-priori assumptions of available behavioral options and of a decision making mechanism, followed by subsequent analysis of evolutionary optimal behavior. For instance, a combat for mating opportunities or territorial possession can be modeled by the War of Attrition game\cite{Bishop1978,Smith1974}: competitors decide whether to retreat or pursue the fighting under time increasing cost for competition. An alternative description is a strategy choice game, such as Hawk-Dove, with options for selfish (Hawk) and cooperative (Dove) behaviors. Evolutionary payoffs for possible outcomes of the contest (retreat vs. pursue, Hawk vs. Dove etc.) define evolutionarily stable strategies (ESS), capable of outperforming any other behavior. In the case of War of Attrition, for instance, evolutionarily stable strategy is to pursue fighting with a probability exponentially decaying with time. The notion of evolutionarily stable behavior stems from game theoretical Nash equilibrium\cite{Nash1950}, though may significantly deviate from it in  interactions containing significant information exchange\cite{Gintis2007}.

Predictions of animal behavior by its evolutionary stability are, in some cases, ambiguous and paradoxical. There is a longstanding argument concerning emergence and maintenance of cooperation or even altruism, which are apparently in discord with the Darwinian ban to increase fitness of others\cite{Wilson2007,Noe2006}. For instance, contrary to the evolutionarily stable strategy for War of Attrition, non-exponential statistics of combat durations takes place in some species. The models, therefore, lack the ability to account for the variety of behavioral strategies in nature.

Back in 1983, S. Austad proposed the combat of male bowl and doily spider ({\it Frontinella pyramitela}) as a quantitative test of evolutionary game theory\cite{Austad1983}. In nature, mature males of this species wander in search of female webs with eggs to fertilize and determine their evolutionary success by fighting with other males. The contests end when one of the combatants gains access to the eggs while his opponent either retreats or dies. The statistics of fight outcomes (percentage of fatal injuries) together with expected evolutionary payoffs (average eggs per female) were accurately documented.

One can expect a self-consistent evolutionary theory predicting spiders' behavior in a given fight from the expected number of eggs per female and from lifetime male reproductive success: mortal combat becomes more reasonable if a future evolutionary gain is unlikely. In laboratory conditions, two virgin spiders of the same size with no previous knowledge of females' value, finish their fight in $67\%$ of cases by death of one of the competitors. The surviving looser gets less than $\delta=5\%$ of eggs. According to field data, average value of a female is $V_{F}=10$ eggs and the male's lifetime reproductive success in the absence of fighting costs is $V_{L}=16.2$ eggs.

The main impediment to derive the spider's observable behavior from the corresponding evolutionary payoffs is the lack of a general theory for biological cognitive processes. Assumption of random choice by a spider whether to retreat or continue the fight leads to a fallacious conclusion that sometimes they do not fight at all. Hypothesis of war of attrition (fight until one of the competitors retreats) predicts much shorter fight durations than observed\cite{Austad1983}. It indicates that even natural seeming assumptions, such as the skill to recognize a retreating competitor, may be beyond the spiders abilities and have to be incorporated in evolutionary models with great caution.

In this work, the complex process of choosing either selfish $S$ (e.g. continue the fight) or cooperative $C$ (e.g. retreat) behavior is described by conditional probabilities $\alpha$ and $\beta$, where $\alpha$ and $1-\alpha$ are respectively the probabilities to choose cooperation $C$ and selfish $S$ behaviors against the competitor's unconditional selfish behavior $S$, while $\beta$ and $1-\beta$ are the probabilities to choose cooperation $C$ and selfish $S$ behaviors against the competitor's unconditional cooperation behavior $C$. In this context, $\alpha$ and $\beta$ can be viewed as selfishness aversion and cooperation attraction respectively. These two parameters suffice to describe all four types of interactions: $S$ vs. $S$, $S$ vs. $C$, $C$ vs. $S$ and $C$ vs. $C$. This approach generalizes previous works based on conditional description of behavior\cite{Nowak1990,Skyrms2002,Bergstrom2003,Kussell2005,Feigel2008} and eliminates any assumptions about specific decision making mechanism.

The conditional probabilities $\alpha$ and $\beta$ can be derived experimentally in two ways. A first, bottom-up approach, is to determine $(\alpha_{observ},\beta_{observ})$ from observable field data $\Omega_{pq}$, the statistics of outcomes $p$ vs. $q$ in the course of a competition $(p,q\in\{S,C\})$ (see Fig. 1). The conditional probability $\alpha_{observ}$ to choose cooperation $C$ against the competitor's unconditional selfish behavior $S$, is the ratio between the unconditional probability $\Omega_{CS}$ (to be cooperative vs. selfish competitor) and the unconditional probability of the competitor to be selfish $\Omega_{CS}+\Omega_{SS}$. Applying the same reasoning for $\beta_{observ}$ one obtains:
\begin{eqnarray}
\alpha_{observ}=\frac{\Omega_{CS}}{{\Omega_{CS}+\Omega_{SS}}},\;\beta_{observ}=\frac{\Omega_{CC}}{{\Omega_{SC}+\Omega_{CC}}}.
\label{alpomega0}%
\end{eqnarray}

A second, top-down, approach derives the evolutionarily stable behavior $(\alpha_{calc},\beta_{calc})$ from observable payoff values  $W_{pq}$, for $p$ vs. $q$ interaction:
\begin{equation}
W_{pq}=\begin{array}{c|c|c}
  & C & S \\
\hline
C & 1 & c \\
\hline
S & b & 0  \\
\end{array}. \label{Wtable0}
\end{equation}
This approach permits to calculate the  $\alpha_{calc}$ and $\beta_{calc}$ parameters only in cases where no symmetrical cooperation $\Omega_{CC}=0$ or symmetrical selfishness $\Omega_{SS}=0$ exist (see Methods). For the case $\Omega_{CC}=0$ one obtains:
\begin{eqnarray}
\alpha_{calc}=\frac{c}{{b}},\;\beta_{calc}=0
\label{stsol1}%
\end{eqnarray}
where for $\Omega_{SS}=0$:
\begin{eqnarray}
\alpha_{calc}=1,\;\beta_{calc}=\frac{b-c}{{1-c}}.
\label{stsol2}%
\end{eqnarray}
No prediction based solely on statistics of interactions can be made for other cases. To be self-consistent, any evolutionary explanation of behavior has to achieve a reasonable match between the values $(\alpha_{observ},\beta_{observ})$ and $(\alpha_{calc},\beta_{calc})$.

In the case of the bowl and doily spider, the observed interaction statistics are $\Omega_{SS}=0.67,\;\Omega_{CC}=0,\;\Omega_{SC}=\Omega_{CS}=0.165$, assuming that only selfish vs. selfish ($67\%$) interactions lead to death of one of the combatants. Following eq. (\ref{alpomega0}), the corresponding observed conditional probabilities are $\alpha_{observ}=0.198$ and $\beta_{observ}=0$. The derived payoff table from $V_{F}=10$, $V_{L}=16.2$ and $\delta=5\%$ is (see Methods):
\begin{equation}
W_{pq}=\begin{array}{c|c|c}
  & C & S \\
\hline
C & 1 & 0.44 \\
\hline
S & 1.55 & 0  \\
\end{array}, \label{Wtable1}
\end{equation}
Following eq. (\ref{stsol1}), the corresponding calculated conditional probabilities are $\alpha_{calc}=0.286$ and $\beta_{calc}=0$ and $\Omega_{SS}=0.55,\;\Omega_{CC}=0,\;\Omega_{SC}=\Omega_{CS}=0.225$ (see eq. (\ref{Momegasyssol1})). An exact match between $(\alpha_{observ},\beta_{observ})$ and $(\alpha_{calc},\beta_{calc})$ is achieved if the lifetime gain value is lowered to $V_{L}=13.42$. This correction seems reasonable since the original estimation of $V_{L}=16.2$ did not take into account naturally existing threats such as predators, etc\cite{Austad1983}.

The analysis of this work is sufficient to derive a self consistent link between evolutionary payoffs $W_{pq}$ and behavior characteristics which depend solely on outcomes statistics $\Omega_{pq}$. For instance, mutual information exchange between competitors $I$ is such a characteristic\cite{Feigel2008}, with a calculated value of $I=0.047$ bits (out of a maximum of $1$ bit describing a binary interaction) (see Methods, eq. (\ref{mutinfo2})). It indicates a low level of communication between fighting males. Only the link between payoffs and $\Omega_{pq}$ depending phenomena may be universal. Other observable parameters, such as duration of fight\cite{Austad1983,Leimar1991}, can not be predicted in the framework of this work and are in general species' specific.

One can derive the payoff table $W_{pq}$ (\ref{Wtable0}) from observable statistics of interaction outcomes $\Omega_{pq}$, even when direct measurement of the payoffs is impossible. When the observed interaction statistics lack symmetrical cooperation or selfish interactions ($\Omega_{CC}=0$ or $\Omega_{SS}=0$) then the values of corresponding payoffs follow from eqs. (\ref{stsol1}) and (\ref{stsol2}) respectively, using $(\alpha_{observ},\beta_{observ})$ (\ref{alpomega0}) instead of $(\alpha_{calc},\beta_{calc})$. But for cases where $\Omega_{CC}\neq 0$ and $\Omega_{SS}\neq 0$, the reconstruction of payoffs from statistics is impossible (see Methods).

Evolutionarily stable behavior $(\alpha_{calc},\beta_{calc})$ follows from known payoffs (\ref{Wtable0}) using eqs. (\ref{stsol1}) and (\ref{stsol2}). The corresponding results are presented in Fig. 2, where payoffs are separated in standard games\cite{Browning2004,Doebeli2005} of Leader $(b>c,c>1)$, Battle of the Sexes $(b<c,b>1)$, Chicken $(b>1,0<c<1)$ and  Prisoner's dilemma $(b>1,c<0)$. Other evolutionarily stable behaviors corresponding to the same payoffs may exist, since a general stability analysis is impossible for behaviors with either $\beta\neq 0$ or $\alpha\neq 1$ (see Methods). The value $\beta=0$, however, is the most intuitive behavior for $c<b$, since it suggests always a selfish behavior versus a cooperative competitor (if $c>b$ then $\alpha=1$ is the complementary case). One can hypothesize, therefore, that the derived behavior $(\alpha_{calc},\beta_{calc})$ is stable and unique in a majority of real world interactions.

Applying this reasoning to the problem of predator inspection by stickleback fish\cite{Milinski1987}, we can argue that sticklebacks do not play the games of Chicken, Leader or Battle of the Sexes, since the observed $\Omega_{CC}=0.207,\;\Omega_{SS}=0.621\;\Omega_{SC}=\Omega_{CS}=0.086$ (leading to $(\alpha_{observ}=0.121$ and $\beta_{observ}=0.708)$, see eq. (\ref{alpomega0})) are not negligible (see Supplementary Material (SM)). It supports indirectly the conclusions that Prisoner's dilemma is the most probable type of the corresponding interaction and highlights the importance of communication in the course of the predator inspection process\cite{Noe2006}. Mutual information exchange of sticklebacks eq. (\ref{mutinfo2}) is $I=0.24$ bits, five times greater than that of the bowl and doily spiders. According to the work, additional information concerning processes of mutual communication must be taken into account to derive a corresponding payoff table.

The developed method allows to analyze the main properties of behavior even in the case when exact statistics of interaction are unknown. For instance, in the case of territory defense by female lions\cite{Heinsohn1995} specific individuals (leaders) cooperatively invest in territory defense, while laggards selfishly avoid risk of fighting. The individual role develops early in life and persists into adulthood. A leader cooperatively takes a significant personal risk even when she recognizes the lagging behavior of others. On a speculative level, one can suggest that the unconditional cooperation stems from development of evolutionarily stable behavior characterized by solitary acts of cooperation $\Omega_{CS},\Omega_{SC}\neq 0$, with subsequent degradation of the ability to change  roles. Such development requires payoffs corresponding to Chicken, Leader or Battle of the Sexes games.

From a game theoretical perspective, the evolutionarily stable behavior conditions (\ref{stsol1}) and (\ref{stsol2}) correspond to a new type of game equilibrium. Whereas the Nash equilibrium\cite{Nash1950} does not consider the possibility of communication between players (thus taking into account only uncorrelated responses (mixed strategies)), and correlated\cite{Aumann1974}/communication\cite{Myerson1986,Forges1986} equilibria consider only fixed information exchange mechanisms, our approach assumes that information exchange mechanisms and behaviors co-evolve, eventually reaching together a stable state. It applies to evolving interactions where the competitors observe each other and exchange information mutually\cite{Noe2006}.

To conclude, evolutionary payoffs are demonstrated to follow unequivocally from observable statistics of interactions lacking symmetrical cooperation or selfishness. The results are verified by using known payoffs and interaction statistics of male combat in bowl and doily spider. In addition, the impossibility to derive payoffs from observations is demonstrated in the case of coexisting acts of mutual cooperation and selfishness. These results may explain solitary acts of cooperation as in the case of territorial defense by lions and support Prisoner's dilemma as the main model for predator inspection by sticklebacks. The correspondence of the derived and observed behaviors supports the possible existence of a unified game theoretical framework for animal behavior\cite{Gintis2007}.

\section{Methods}

\subsection{Evolutionarily stable points}

To derive eqs. (\ref{stsol1}) and (\ref{stsol2}), one should find for each payoff table $W_{pq}$ an evolutionarily stable behavior $(\alpha_{st},\beta_{st})$, that outperforms any other behavior (mutant) $m$, characterized by $(\alpha_{m},\beta_{m})$:
\begin{equation}
G(\alpha_{st},\beta_{st}|\alpha_{st},\beta_{st})>G(\alpha_{m},\beta_{m}|\alpha_{st},\beta_{st}), \label{MG1}%
\end{equation}
where $G(\alpha_{i},\beta_{i}|\alpha_{j},\beta_{j})$ is the gain of an individual $(\alpha_{i},\beta_{i})$ interacting with $(\alpha_{j},\beta_{j})$. On the boundary of the $(\alpha,\beta)$ space ($\alpha=0$, $\alpha=1$, $\beta=0$ or $\beta=1$), the stable values $(\alpha_{st},\beta_{st})$ correspond to the maxima of the gain in the tangential direction to the boundary:
\begin{eqnarray}
\left .\frac{\partial G(\alpha_{i},\beta_{i}|\alpha_{j},\beta_{j})}{{\partial x_{i}}}\right |_{\substack{\alpha_{i,j}=\alpha_{st} \\
\beta_{i,j}=\beta_{st}}}=0,\label{Mstcond1}
\end{eqnarray}
\begin{eqnarray}
\left .\frac{\partial^{2} G(\alpha_{i},\beta_{i}|\alpha_{j},\beta_{j})}{{\partial x_{i}^{2}}}\right |_{\substack{\alpha_{i,j}=\alpha_{st} \\
\beta_{i,j}=\beta_{st}}}<0,
\label{Mstcond2}%
\end{eqnarray}
where $x_{i}$ denotes either $\alpha_{i}$ or $\beta_{i}$ (e.g. if $\beta=0$ then the tangential direction $x_{i}$ is $\alpha_{i}$). Mutants along perpendicular direction do not affect the values $(\alpha_{st},\beta_{st})$ though may make the corresponding behavior to be unstable (see SM, Fig. 3).

The gain $G(\alpha_{i},\beta_{i}|\alpha_{j},\beta_{j})$ is determined by the probabilities $\Omega_{pq}$ of an interaction $p$ vs. $q$, and by the corresponding payoffs $W_{pq}$, $G=\sum\Omega_{pq}W_{pq}$:
\begin{eqnarray}
G(\alpha_{i},\beta_{i}|\alpha_{j},\beta_{j})=\Omega_{CC}+c\Omega_{CS}+b\Omega_{SC},
\label{MG2}%
\end{eqnarray}
taking eq. (\ref{Wtable0}) into account.

In the course of an interaction between individuals $(\alpha_{i},\beta_{i})$ and $(\alpha_{j},\beta_{j})$, the probabilities $\Omega_{pq}$ are determined by a system of equations analogous to eq. (\ref{alpomega0})):
\begin{eqnarray}
&&\alpha_{i}=\frac{\Omega_{CS}}{{\Omega_{CS}+\Omega_{SS}}},\;\alpha_{j}=\frac{\Omega_{SC}}{{\Omega_{SS}+\Omega_{SC}}},\nonumber \\
&&\beta_{i}=\frac{\Omega_{CC}}{{\Omega_{SC}+\Omega_{CC}}},\;\beta_{j}=\frac{\Omega_{CC}}{{\Omega_{CC}+\Omega_{CS}}},\nonumber \\
&&\Omega_{SS}+\Omega_{SC}+\Omega_{CS}+\Omega_{CC}=1. \label{Momegasys}%
\end{eqnarray}
This system, comprising $5$ equations for $4$ variables, is solvable for five distinct cases only:  if the competitors are identical and on the boundaries of $(\alpha,\beta)$ space. The corresponding solution for identical competitors $(\alpha_{i},\beta_{i})=(\alpha_{j},\beta_{j})=(\alpha,\beta)$ is:
\begin{eqnarray}
\Omega_{CC}&=&\frac{\alpha\beta}{{1-\beta+\alpha}},\;\Omega_{SS}=\frac{(1-\alpha)(1-\beta)}{{1-\beta+\alpha}},\nonumber \\
\Omega_{CS}&=&\Omega_{SC}=\frac{\alpha(1-\beta)}{{1-\beta+\alpha}},\nonumber \\
\label{Momegasyssol0}%
\end{eqnarray}
Second, if they are on the $\beta=0$ boundary $(\beta_{i}=\beta_{j}=0)$:
\begin{eqnarray}
\Omega_{CC}&=&0,\;\Omega_{SS}=\frac{(1-\alpha_{i})(1-\alpha_{j})}{{1-\alpha_{i}\alpha_{j}}},\nonumber \\
\Omega_{CS}&=&\frac{\alpha_{i}(1-\alpha_{j})}{{1-\alpha_{i}\alpha_{j}}},\;\Omega_{SC}=\frac{\alpha_{j}(1-\alpha_{i})}{{1-\alpha_{i}\alpha_{j}}},\nonumber \\
\label{Momegasyssol1}%
\end{eqnarray}
Third, if they are on the $\alpha=1$ boundary $(\alpha_{i}=\alpha_{j}=1)$:
\begin{eqnarray}
\Omega_{CC}&=&\frac{\beta_{i}\beta_{j}}{{\beta_{i}+\beta_{j}-\beta_{i}\beta_{j}}},\;\Omega_{SS}=0,\nonumber \\
\Omega_{CS}&=&\frac{\beta_{i}(1-\beta_{j})}{{\beta_{i}+\beta_{j}-\beta_{i}\beta_{j}}},\;\Omega_{SC}=\frac{\beta_{j}(1-\beta_{i})}{{\beta_{i}+\beta_{j}-\beta_{i}\beta_{j}}},\nonumber \\
\label{Momegasyssol2}%
\end{eqnarray}
Fourth, $(\alpha_{i}=\alpha_{j}=0)$: $\Omega_{SS}=1,\;\Omega_{SC}=\Omega_{CS}=\Omega_{CC}=0$. Fifth, $(\beta_{i}=\beta_{j}=1)$: $\Omega_{CC}=1,\;\Omega_{SC}=\Omega_{CS}=\Omega_{SS}=0$.

Eqs. (\ref{stsol1}) and (\ref{stsol2}) for $(\alpha_{st},\beta_{st})$ follow from eqs. (\ref{Mstcond1}) and (\ref{MG2}) using (\ref{Momegasyssol1}) and (\ref{Momegasyssol2}) respectively. This derivation considerers the mutants along the boundaries only (see SM, Fig. 3). No specific $(\alpha_{st},\beta_{st})$ exists in the fourth and fifth cases ($\alpha=0$ and $\beta=1$ boundaries), since the corresponding $\Omega_{pq}$ are independent of $\alpha$ and $\beta$.

A general stability analysis of particular behavior $(\alpha,\beta)$ is impossible. It requires analysis of host vs. mutant interactions that can not be described by conditional probabilities $(\alpha,\beta)$. For instance, when mutants leave the boundary, the system of eqs. (\ref{Momegasys}) is unsolvable. In case of the bowl and doily spider the stability of behavior at $(\alpha=0.286,\beta=0)$ along the $\beta$ axis is justified mainly by observation (see SM, Fig. 3). Stability of behavior along $\alpha$ axis can be justified only by simulation of population dynamics (see SM), since the stability condition (\ref{Mstcond2}) vanishes for the gain (\ref{MG2}) in the cases described by statistics (\ref{Momegasyssol1}) and (\ref{Momegasyssol2}), with behaviors (\ref{stsol1}) and (\ref{stsol2}) respectively.

\subsection{Reduction of payoff table $W_{pq}$ to two parameter $b$ and $c$ form}

The payoffs of biological organisms competing for a resource with value $V_{F}$, and possessing lifetime gain $V_{L}$ are:
\begin{equation}
\begin{array}{c|c|c}
  & C & S \\
\hline
C & W_{CC} & W_{CS} \\
\hline
S & W_{SC} & W_{SS}  \\
\end{array}=\begin{array}{c|c|c}
  & C & S \\
\hline
C & V_{F}/2 & V_{F}\delta \\
\hline
S & V_{F}(1-\delta) & V_{F}/2-V_{L}/2  \\
\end{array}, \label{MWtable2}
\end{equation}
where each of them receives $V_{F}/2$ in case of mutual cooperation, $V_{F}/2-V_{L}/2$ for mutual selfishness (assuming that it leads to the death of one of the combatants), while $V_{F}\delta$ and $V_{F}(1-\delta)$ are the shares of interacting selfish and cooperative competitors respectively.

Reduction of the payoff table (\ref{MWtable2}) to its two parameters form (\ref{Wtable0}) requires two transformations:
\begin{equation}
W'_{pq}=W_{pq}-W_{SS}, \label{Mtran1}
\end{equation}
and
\begin{equation}
W''_{pq}=W'_{pq}/(W'_{CC}-W'_{SS}). \label{Mtran2}
\end{equation}
Consequently, the parameters $b$ and $c$ in (\ref{Wtable0}) are:
\begin{eqnarray}
b&=&\frac{W_{SC}-W_{SS}}{{W_{CC}-W_{SS}}},\nonumber \\
c&=&\frac{W_{CS}-W_{SS}}{{W_{CC}-W_{SS}}}. \label{Mbc}
\end{eqnarray}
These transformation do no affect the stability conditions (\ref{Mstcond1}) and (\ref{Mstcond2}) (see SM).

Eq. (\ref{Wtable1}) follows substituting $V_{F}=10$, $V_{L}=16.2$ and $\delta=5\%$ to eqs. (\ref{MWtable2} and (\ref{Mbc}).

\subsection{Information calculation}

Mutual information per interaction is given by\cite{Feigel2008}:
\begin{eqnarray}
I(\alpha,\beta)=\log_{2}\left (\frac{(1-\alpha)^{(1-\alpha)\gamma}(1-\beta)^{(1-\beta)(1-\gamma)}\alpha^{\alpha\gamma}\beta^{\beta(1-\gamma)}}{{\gamma^{\gamma}(1-\gamma)^{\gamma}}}\right ),\nonumber \\
\label{mutinfo2}
\end{eqnarray}
where:
\begin{equation}
\gamma=\frac{1-\beta}{{1+\alpha-\beta}}. \label{gammaii}%
\end{equation}

For spiders $I=0.047$ bits, substituting $(\alpha=0.198,\beta=0)$ in eq. (\ref{mutinfo2}). The same result for sticklebacks is  $I=0.24$ bits, $(\alpha=0.121,\beta=0.708)$.

\section{Supplementary materials}

\subsection{Solution of system (\ref{Momegasys})}

The system of eqs. (\ref{Momegasys}) can be represented as:
\begin{equation}
\left(
  \begin{array}{ccccc}
    \alpha_{1} & 0 & \alpha_{1}-1 & 0 \\
    \alpha_{2} & \alpha_{2}-1 & 0 & 0 \\
    0 & \beta_{1} & 0 & \beta_{1}-1 \\
    0 & 0 & \beta_{2} & \beta_{2}-1 \\
    1 & 1 & 1 & 1  \\
  \end{array}
\right)
\left(
  \begin{array}{c}
    \Omega_{SS} \\
    \Omega_{SC} \\
    \Omega_{CS} \\
    \Omega_{CC} \\
  \end{array}
\right)=
\left(
  \begin{array}{c}
    0 \\
    0 \\
    0 \\
    0 \\
    1 \\
  \end{array}
\right).\label{SMomegasysmat}
\end{equation}
A solution for this system exists if the determinant of the corresponding extended matrix $U$ vanishes:
\begin{equation}
\det U = 0,\label{SMdetU0}
\end{equation}
where
\begin{equation}
U = \left(
  \begin{array}{ccccc}
    \alpha_{1} & 0 & \alpha_{1}-1 & 0 & 0 \\
    \alpha_{2} & \alpha_{2}-1 & 0 & 0 & 0 \\
    0 & \beta_{1} & 0 & \beta_{1}-1 & 0 \\
    0 & 0 & \beta_{2} & \beta_{2}-1 & 0 \\
    1 & 1 & 1 & 1 & 1 \\
  \end{array}
\right).\label{SMUmat}
\end{equation}
For a general case there is no solution, since it includes five equations and only four variables.

The determinant of $U$:
\begin{eqnarray}
&&\det U(\alpha_{1},\beta_{1},\alpha_{2},\beta_{2}) = \nonumber \\
&&\alpha_{1}(\beta_{1}-1)\beta_{2}+\alpha_{2}(\alpha_{1}\beta_{2}-\beta_{1}(\alpha_{1}+\beta_{2}-1)).\nonumber \\
\label{SMdeter1}
\end{eqnarray}
vanishes either in case of identical phenotypes $(\alpha_{1},\beta_{1})=(\alpha_{2},\beta_{2})$ or at the boundaries of the $(\alpha,\beta)$ space, $\alpha=0,1$ or $\beta=0,1$. It can be demonstrated as follows:
\begin{equation}
\det U(\alpha,\beta,\alpha,\beta) = 0,\label{SMdetUeq}
\end{equation}
Then taking into account (\ref{SMdetUeq}), $\alpha_{2}\neq\alpha_{1}$ fits (\ref{SMdetU0}) if:
\begin{equation}
\left .\frac{\partial\det U}{{\partial \alpha_{2}}}\right |_{\alpha_{1,2}=\alpha, \beta_{1,2}=\beta} = \beta(1-\beta),\label{SMdetUderalpha}
\end{equation}
vanishes. The same holds for changes in $\beta_{2}$:
\begin{equation}
\left .\frac{\partial\det U}{{\partial \beta_{2}}}\right |_{\alpha_{1,2}=\alpha, \beta_{1,2}=\beta} = -\alpha(1-\alpha).\label{SMdetUderbeta}
\end{equation}

The solutions (\ref{Momegasyssol0}), (\ref{Momegasyssol1}) and (\ref{Momegasyssol2}) can be checked by a substitution.

\subsection{Normalization}

The transformations (\ref{Mtran1}) and (\ref{Mtran2}) do not affect the stability condition (\ref{MG1}). Taking into account that $G=\sum\Omega_{pq}W_{pq}$, (\ref{MG1}) becomes $\sum_{pq}\Delta\Omega_{pq}W_{pq}\leq 0$:
\begin{eqnarray}
&&\Delta\Omega_{CC}W_{CC}+\Delta\Omega_{CS}W_{CS}+\ldots\nonumber \\
&+&\Delta\Omega_{SC}W_{SC}+\Delta\Omega_{SS}W_{SS}\leq 0. \label{DO10}%
\end{eqnarray}
where $\Delta\Omega_{pq}=\left (\Omega^{m}_{pq}-\Omega^{st}_{pq}\right )$.

Applying the first transformation (\ref{Mtran1}) to (\ref{DO10}) results in:
\begin{eqnarray}
&&\Delta\Omega_{CC}W'_{CC}+\Delta\Omega_{CS}W'_{CS}+\ldots\nonumber \\
&+&\Delta\Omega_{SC}W'_{SC}+\Delta\Omega_{SS}W'_{SS}+\ldots\nonumber \\
&+&W_{SS}\sum_{pq}\Delta\Omega_{pq}\leq 0. \label{b1GB}%
\end{eqnarray}
The last term in the left part vanishes $(\sum_{pq}\Delta\Omega_{pq}=0)$ preserving the form of condition (\ref{MG1}). The second transformation (\ref{Mtran2}) converts (\ref{b1GB}) into $\sum_{pq}\Delta\Omega_{pq}W''_{pq}\leq 0$.

\subsection{Stability analysis}

Evolutionary stability of $(\alpha_{st},\beta_{st})$ (a solution of (\ref{Mstcond1})) can be justified by population dynamics analysis, if the corresponding condition (\ref{Mstcond2}) vanishes. The population dynamics is described by the replicator dynamics equations \citep{Taylor1978}:
\begin{equation}
\frac{\partial\rho(\alpha_{i},\beta_{i})}{{\partial t}}=\frac{1}{{T_{gen}}}\rho(\alpha_{i},\beta_{i})\frac{F(\alpha_{i},\beta_{i})-\overline F}{{|\overline F|}}, \label{SMrepdyn}%
\end{equation}
where $\rho(\alpha_{i},\beta_{i})$ the density of the behavior $(\alpha_{i},\beta_{i})$, $T_{gen}$ is the time span of a single generation and $F(\alpha_{i},\beta_{i})=\sum_{j} \rho(\alpha_{j},\beta_{j})G(\alpha_{i},\beta_{i}|\alpha_{j},\beta_{j})$
is the evolutionary fitness (average gain) of the phenotype $(\alpha_{i},\beta_{i})$. The average fitness $\overline F$ in the population is
$\overline F= \sum_{i}\rho(\alpha_{i},\beta_{i})F(\alpha_{i},\beta_{i})$.

In the case of bowl and doily spiders, convergence of a population to $\alpha=0.286$ (corresponding to condition (\ref{stsol1}) in case of payoffs $(b=1.556,c=0.444)$) is checked by a simulation of replicator dynamics eq. (\ref{SMrepdyn}) taking into account gain (\ref{MG2}), statistics (\ref{Momegasyssol1}) and prediction (\ref{stsol1}) (see the attached animation). The same convergence holds for all solutions (\ref{stsol1}) with payoffs $(b,c)$ corresponding to games of Chicken $(b>1,0<c<1)$ and Leader $(b>c,c>1)$.

The stability of $\beta$ values corresponding to condition (\ref{stsol2}) can be verified for the entire region of Battle of the Sexes game $(b<c,b>1)$ using gain (\ref{MG2}) with statistics (\ref{Momegasyssol2}).

\subsection{Calculation of $\Omega_{pq}$, $\alpha$ and $\beta$ in case of predator inspection by sticklebacks}

M. Milinski checked the readiness of a single stickleback fish to inspect a predator (located in the front of a pool) in presence of permanently cooperative or defecting companion, which was simulated by a tilted mirror\cite{Milinski1987}. The "cooperating mirror" was parallel to the pool, simulating a companion which followed immediately a proceeding stickleback. The "defecting mirror" was tilted to simulate a companion which stayed increasingly behind a proceeding individual. The results which are relevant for this work are\cite{Milinski1987}: "With the cooperating mirror the sticklebacks twice as often in the front half (mean $P_{C}=25.1\%$) as with the defecting mirror (mean $P_{S}=12.1\%$)", where $P_{C}$ and $P_{S}$ are the probabilities to observe a stickleback close to the predator in the experiments with cooperative and defecting mirrors respectively.

Associating predator inspection (going to the front half of the pool) with cooperative behavior and taking into account that in the case of a "cooperating mirror" only $S\;vs.\;S$ or $C\;vs.\;C$ interactions are observable, one obtains:
\begin{eqnarray}
\frac{\Omega_{CC}}{{\Omega_{SS}+\Omega_{CC}}}=P_{C}. \label{SMmil1}
\end{eqnarray}
The same applied to the case of "defecting mirror" ($C\;vs.\;S$ or $S\;vs.\;S$ interactions are observable) leads to:
\begin{eqnarray}
\frac{\Omega_{CS}}{{\Omega_{SS}+\Omega_{CS}}}=P_{S}, \label{SMmil2}
\end{eqnarray}
where $\Omega_{pq}$ are the probabilities of the outcomes in a hypothetical experiment with two sticklebacks (rather than a stickleback and a mirror) possessing the same $(\alpha,\beta)$.

To derive conditional probabilities $(\alpha,\beta)$ of a stickleback  and corresponding interaction statistics $\Omega_{pq}$,  substitution of $P_{C}=0.251$ and $P_{S}=0.121$ to eqs. (\ref{SMmil1}), (\ref{SMmil2}) and (\ref{Momegasyssol0}) results in $\alpha=P_{S}=0.121$ and $\beta=(1-\alpha)/(1+2\alpha)=0.708$. The corresponding values $\Omega_{pq}$ following (\ref{Momegasyssol0}) are $\Omega_{CC}=0.207$, $\Omega_{SS}=0.621$ and $\Omega_{SC}=\Omega_{CS}=0.086$. As one can see, neither $\Omega_{CC}$ nor $\Omega_{SS}$ are negligible in this case.

\subsection{Lions defending their territory}

Female lions defend the pride's territory by approaching a potential intruder and when necessary proceeding to a fight\cite{Heinsohn1995}. To avoid risk of fighting "certain individuals consistently lag behind their companions during the group response". The most surprising phenomenon is that "although leaders recognize laggards and behave more cautiously in their presence, they continue to lead the response". The separation into selfish laggards and cooperative leaders "appears early in life and persists into adulthood".

In this case the reported observations are insufficient to derive statistics of interaction outcomes $\Omega_{pq}$, where laggard and leader are associated with selfish and cooperative behaviors respectively.

\begin{widetext}
\clearpage
\begin{figure}
  \begin{center}
      \resizebox{0.5\textwidth}{!}{\includegraphics{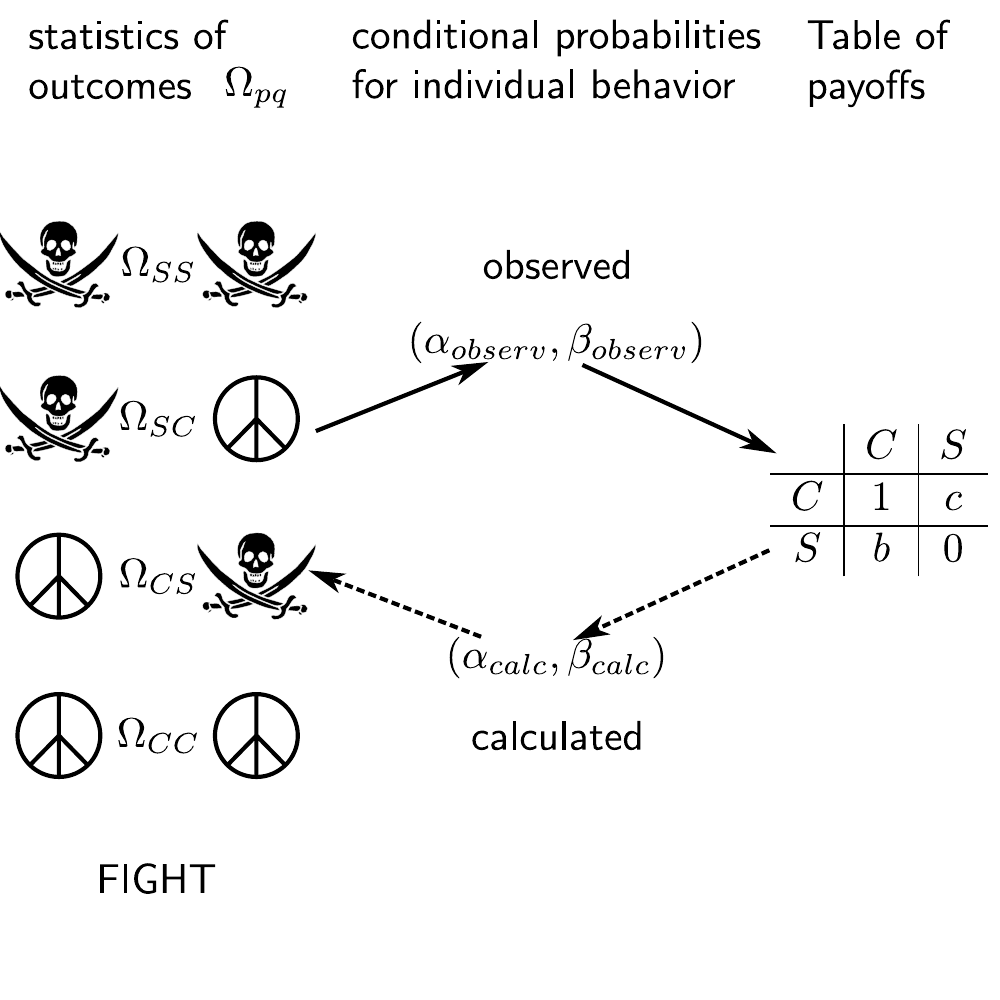}}
    \caption{{\bf From observed behavior to evolutionary payoffs and back.} Animal interactions may be classified according to the behavior of the competitors, using observable parameters such as level of aggression or cooperation. Four types of contests exist when the individual behavior is either selfish $S$ or cooperative $C$ ($S$ vs. $S$, $S$ vs. $C$, $C$ vs. $S$ and $C$ vs. $C$). Continuous observations of such competitions provide $\Omega_{pq}$, the probabilities of $p$ vs. $q$ interactions $(p,q\in \{S,C\})$. The statistics $\Omega_{pq}$ defines individual behavior parameters $\alpha_{observ}$ and $\beta_{observ}$, the conditional probabilities to be cooperative against selfish and cooperative competitors respectively. The corresponding Darwinian payoffs may be determined by analysis of evolutionary stability of the observed behavior. When payoffs are known, one can predict evolutionarily stable behavior $(\alpha_{calc},\beta_{calc})$ and compare it with the observed one $(\alpha_{observ},\beta_{observ})$. Animals with documented statistics of interactions and payoffs, such as the bowl and doily spiders, provide an ultimate test for evolutionary theory.}
    \label{fig1}
  \end{center}
\end{figure}
\clearpage
\begin{figure}
  \begin{center}
      \resizebox{0.5\textwidth}{!}{\includegraphics{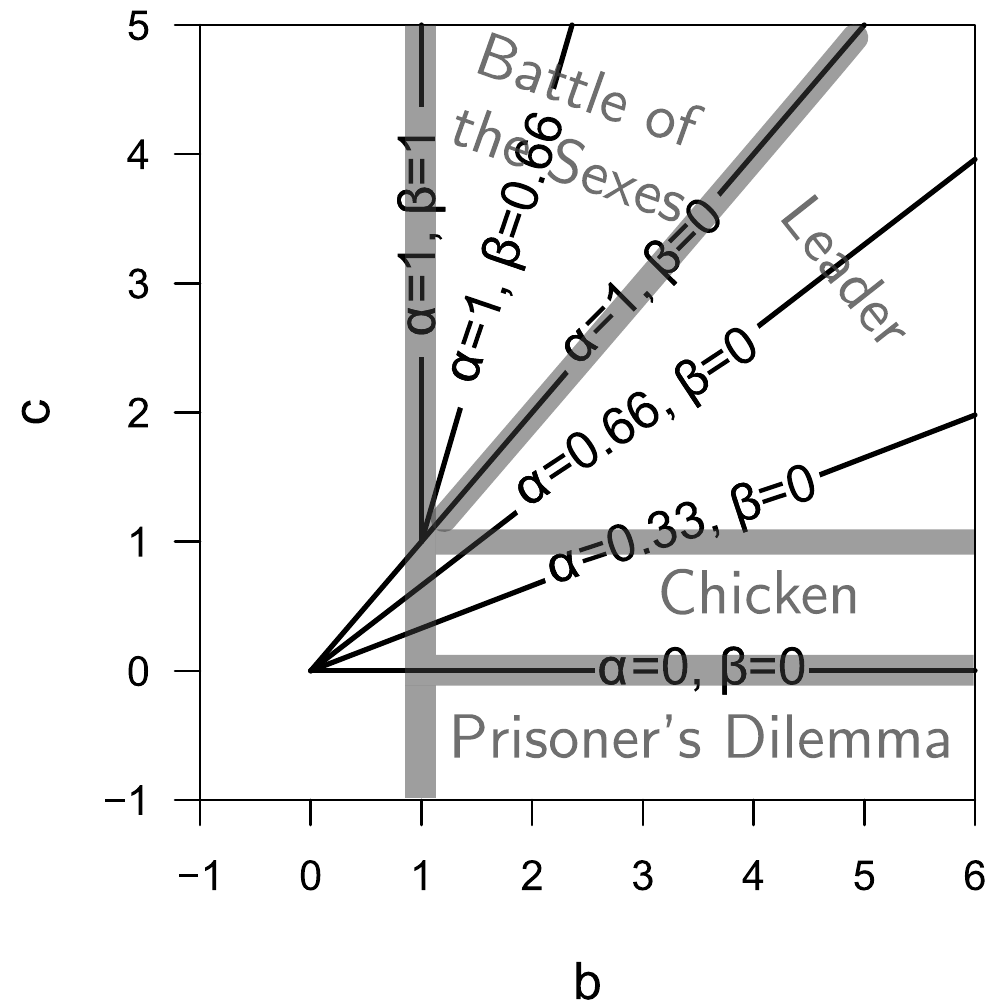}}
    \caption{{\bf Darwinian payoffs and corresponding evolutionarily stable behavior.} Darwinian payoffs $b$ and $c$ (\ref{Wtable0}) are generally presented as model games of Leader $(b>c,c>1)$, Battle of the Sexes $(b<c,b>1)$, Chicken $(b>1,0<c<1)$ and  Prisoner's dilemma $(b>1,c<0)$.  For each values $b$ and $c$ the corresponding prediction of evolutionarily stable behavior $(\alpha,\beta)$ is indicated. Chicken and Leader games cause unconditional selfishness against cooperative competitors ($\beta=0$, with corresponding $\Omega_{CC}=0$). Battle of the Sexes leads to permanent cooperation against selfishness ($\alpha=1$, with corresponding $\Omega_{SS}=0$). No predictions can be made for other cases, since stability analysis of the corresponding $(\alpha,\beta)$ parameters requires previous knowledge of some difficult to observe interaction characteristics, such as information exchange, the order of mutual responses etc.}
    \label{fig2}
  \end{center}
\end{figure}
\clearpage
\begin{figure}
  \begin{center}
      \resizebox{0.5\textwidth}{!}{\includegraphics{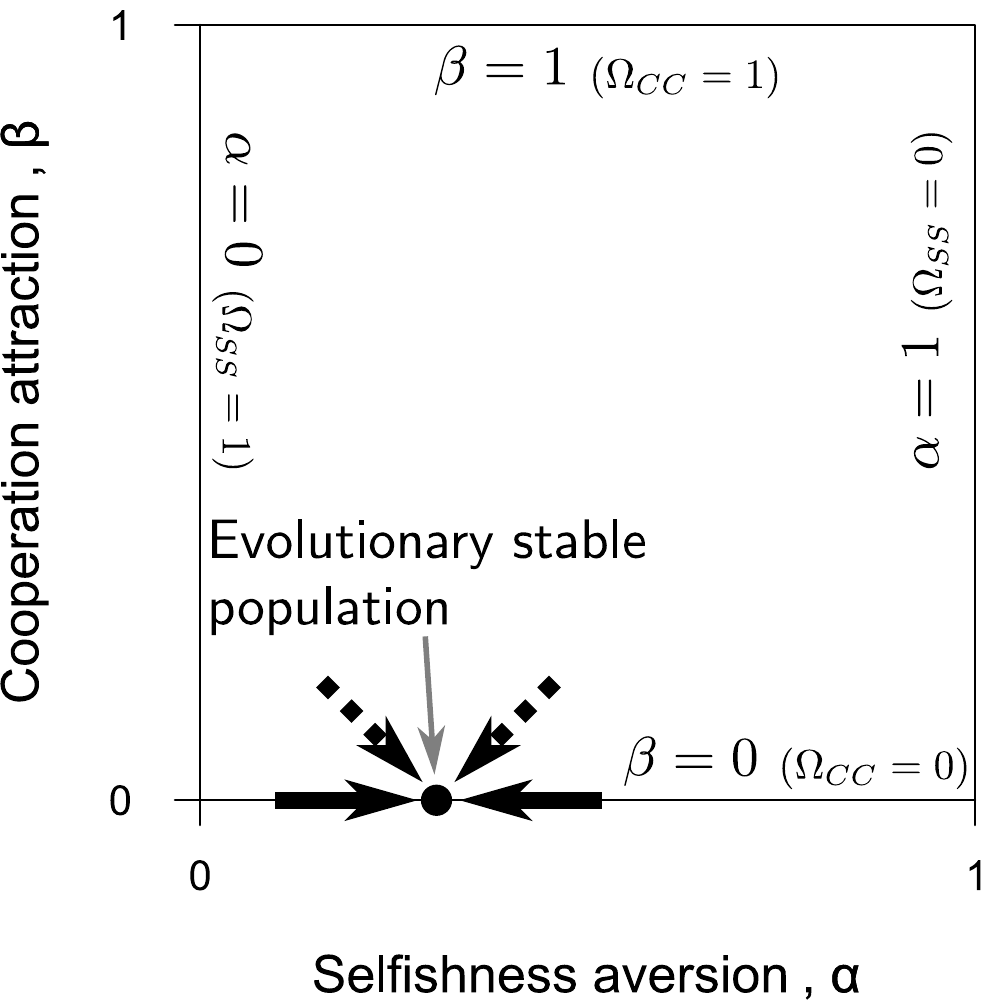}}
    \caption{{\bf Evolutionary stability of behavior $(\alpha,\beta)$.} Evolutionarily stable behavior $(\alpha,\beta)$ has to outperform any possible mutant. In the $(\alpha,\beta)$ space only interactions between individuals on the same boundary can be analyzed ($\beta=0$ with corresponding $\Omega_{CC}=0$, $\alpha=1$ with $\Omega_{SS}=0$, $\beta=1$ with $\Omega_{CC}=1$, $\alpha=0$ with $\Omega_{SS}=1$). Convergence of a population towards its stable position can be analyzed only along the boundaries (solid arrows). Mutants along the boundary will converge towards the stable point, while dynamics of mutants outside the boundary (dashed arrows) can not be generally analyzed. These mutants do not affect the position of the stable point on the boundary.}
    \label{fig3}
  \end{center}
\end{figure}
\end{widetext}

\begin{thebibliography}{10}
\expandafter\ifx\csname url\endcsname\relax
  \def\url#1{\texttt{#1}}\fi
\expandafter\ifx\csname urlprefix\endcsname\relax\def\urlprefix{URL }\fi
\providecommand{\bibinfo}[2]{#2}
\providecommand{\eprint}[2][]{\url{#2}}

\bibitem{Smith1982}
\bibinfo{author}{Smith, J.~M.}
\newblock \emph{\bibinfo{title}{Evolution and the Theory of Games}}
  (\bibinfo{publisher}{Cambridge University Press}, \bibinfo{year}{1982}).

\bibitem{Vincent2005}
\bibinfo{author}{Vincent, T.~L.} \& \bibinfo{author}{Brown, J.~S.}
\newblock \emph{\bibinfo{title}{Evolutionary Game Theory, Natural Selection,
  and Darwinian Dynamics}} (\bibinfo{publisher}{Cambridge University Press},
  \bibinfo{year}{2005}).

\bibitem{Nowak2004}
\bibinfo{author}{Nowak, M.~A.} \& \bibinfo{author}{Sigmund, K.}
\newblock \bibinfo{title}{Evolutionary dynamics of biological games}.
\newblock \emph{\bibinfo{journal}{Science}} \textbf{\bibinfo{volume}{303}},
  \bibinfo{pages}{793 -- 799} (\bibinfo{year}{2004}).

\bibitem{Bishop1978}
\bibinfo{author}{Bishop, D.~T.} \& \bibinfo{author}{Cannings, C.}
\newblock \bibinfo{title}{Generalized war of attrition}.
\newblock \emph{\bibinfo{journal}{J. Theor. Biol.}}
  \textbf{\bibinfo{volume}{70}}, \bibinfo{pages}{85 -- 124}
  (\bibinfo{year}{1978}).

\bibitem{Smith1974}
\bibinfo{author}{Smith, J.~M.}
\newblock \bibinfo{title}{Theory of games and evolution of animal conflicts}.
\newblock \emph{\bibinfo{journal}{J. Theor. Biol.}}
  \textbf{\bibinfo{volume}{47}}, \bibinfo{pages}{209 -- 221}
  (\bibinfo{year}{1974}).

\bibitem{Nash1950}
\bibinfo{author}{Nash, J.}
\newblock \bibinfo{title}{Equilibrium points in n-person games}.
\newblock \emph{\bibinfo{journal}{Proc. Natl. Acad. Sci. USA}}
  \textbf{\bibinfo{volume}{36}}, \bibinfo{pages}{48 -- 49}
  (\bibinfo{year}{1950}).

\bibitem{Gintis2007}
\bibinfo{author}{Gintis, H.}
\newblock \bibinfo{title}{A framework for the unification of the behavioral
  sciences}.
\newblock \emph{\bibinfo{journal}{Behav. Brain Sci.}}
  \textbf{\bibinfo{volume}{30}}, \bibinfo{pages}{1 -- 61}
  (\bibinfo{year}{2007}).

\bibitem{Wilson2007}
\bibinfo{author}{Wilson, D.~S.} \& \bibinfo{author}{Wilson, E.~O.}
\newblock \bibinfo{title}{Rethinking the theoretical foundation of
  sociobiology}.
\newblock \emph{\bibinfo{journal}{Q. Rev. Biol.}}
  \textbf{\bibinfo{volume}{82}}, \bibinfo{pages}{327 -- 348}
  (\bibinfo{year}{2007}).

\bibitem{Noe2006}
\bibinfo{author}{Noe, R.}
\newblock \bibinfo{title}{Cooperation experiments: coordination through
  communication versus acting apart together}.
\newblock \emph{\bibinfo{journal}{Anim. Behav.}} \textbf{\bibinfo{volume}{71}},
  \bibinfo{pages}{1 -- 18} (\bibinfo{year}{2006}).

\bibitem{Austad1983}
\bibinfo{author}{Austad, S.~N.}
\newblock \bibinfo{title}{A game theoretical interpretation of male combat in
  the bowl and doily spider (frontinella-pyramitela)}.
\newblock \emph{\bibinfo{journal}{Anim. Behav.}} \textbf{\bibinfo{volume}{31}},
  \bibinfo{pages}{59 -- 73} (\bibinfo{year}{1983}).

\bibitem{Nowak1990}
\bibinfo{author}{Nowak, M.} \& \bibinfo{author}{Sigmund, K.}
\newblock \bibinfo{title}{The evolution of stochastic strategies in the
  prisoners-dilemma}.
\newblock \emph{\bibinfo{journal}{Acta Appl. Math.}}
  \textbf{\bibinfo{volume}{20}}, \bibinfo{pages}{247 -- 265}
  (\bibinfo{year}{1990}).

\bibitem{Skyrms2002}
\bibinfo{author}{Skyrms, B.}
\newblock \bibinfo{title}{Altruism, inclusive fitness, and "the logic of
  decision"}.
\newblock \emph{\bibinfo{journal}{Philos. Sci.}} \textbf{\bibinfo{volume}{69}},
  \bibinfo{pages}{S104 -- S111} (\bibinfo{year}{2002}).

\bibitem{Bergstrom2003}
\bibinfo{author}{Bergstrom, T.}
\newblock \bibinfo{title}{The algebra of assortative encounters and the
  evolution of cooperation}.
\newblock \emph{\bibinfo{journal}{Int. Game Theor. Rev.}}
  \textbf{\bibinfo{volume}{5}}, \bibinfo{pages}{211 -- 228}
  (\bibinfo{year}{2003}).

\bibitem{Kussell2005}
\bibinfo{author}{Kussell, E.} \& \bibinfo{author}{Leibler, S.}
\newblock \bibinfo{title}{Phenotypic diversity,population growth,and
  information in fluctuating environments}.
\newblock \emph{\bibinfo{journal}{Science}} \textbf{\bibinfo{volume}{309}},
  \bibinfo{pages}{2075 -- 2078} (\bibinfo{year}{2005}).

\bibitem{Feigel2008}
\bibinfo{author}{Feigel, A.}
\newblock \bibinfo{title}{Essential conditions for evolution of communication
  within a species}.
\newblock \emph{\bibinfo{journal}{J. Theor. Biol.}} \textbf{\bibinfo{volume}{In
  Press.}} (\bibinfo{year}{2008}).

\bibitem{Leimar1991}
\bibinfo{author}{Leimar, O.}, \bibinfo{author}{Austad, S.} \&
  \bibinfo{author}{Enquist, M.}
\newblock \bibinfo{title}{A test of the sequential assessment game - fighting
  in the bowl and doily spider frontinella-pyramitela}.
\newblock \emph{\bibinfo{journal}{Evolution}} \textbf{\bibinfo{volume}{45}},
  \bibinfo{pages}{862 -- 874} (\bibinfo{year}{1991}).

\bibitem{Browning2004}
\bibinfo{author}{Browning, L.} \& \bibinfo{author}{Colman, A.~M.}
\newblock \bibinfo{title}{Evolution of coordinated alternating reciprocity in
  repeated dyadic games}.
\newblock \emph{\bibinfo{journal}{J. Theor. Biol.}}
  \textbf{\bibinfo{volume}{229}}, \bibinfo{pages}{549 -- 557}
  (\bibinfo{year}{2004}).

\bibitem{Doebeli2005}
\bibinfo{author}{Doebeli, M.} \& \bibinfo{author}{Hauert, C.}
\newblock \bibinfo{title}{Models of cooperation based on the prisoner's dilemma
  and the snowdrift game}.
\newblock \emph{\bibinfo{journal}{Ecol. Lett.}} \textbf{\bibinfo{volume}{8}},
  \bibinfo{pages}{748 -- 766} (\bibinfo{year}{2005}).

\bibitem{Milinski1987}
\bibinfo{author}{Milinski, M.}
\newblock \bibinfo{title}{Tit-for-tat in sticklebacks and the evolution of
  cooperation}.
\newblock \emph{\bibinfo{journal}{Nature}} \textbf{\bibinfo{volume}{325}},
  \bibinfo{pages}{433 -- 435} (\bibinfo{year}{1987}).

\bibitem{Heinsohn1995}
\bibinfo{author}{Heinsohn, R.} \& \bibinfo{author}{Packer, C.}
\newblock \bibinfo{title}{Complex cooperative strategies in group-territorial
  african lions}.
\newblock \emph{\bibinfo{journal}{Science}} \textbf{\bibinfo{volume}{269}},
  \bibinfo{pages}{1260 -- 1262} (\bibinfo{year}{1995}).

\bibitem{Aumann1974}
\bibinfo{author}{Aumann, R.}
\newblock \bibinfo{title}{Subjectivity and correlation in randomized
  strategies}.
\newblock \emph{\bibinfo{journal}{J. Math. Econ.}}
  \textbf{\bibinfo{volume}{1}}, \bibinfo{pages}{67 -- 96}
  (\bibinfo{year}{1974}).

\bibitem{Myerson1986}
\bibinfo{author}{Myerson, R.~B.}
\newblock \bibinfo{title}{Multistage games with communication}.
\newblock \emph{\bibinfo{journal}{Econometrica}} \textbf{\bibinfo{volume}{54}},
  \bibinfo{pages}{323 -- 358} (\bibinfo{year}{1986}).

\bibitem{Forges1986}
\bibinfo{author}{Forges, F.}
\newblock \bibinfo{title}{An approach to communication equilibria}.
\newblock \emph{\bibinfo{journal}{Econometrica}} \textbf{\bibinfo{volume}{54}},
  \bibinfo{pages}{1375 -- 1385} (\bibinfo{year}{1986}).

\bibitem{Taylor1978}
\bibinfo{author}{Taylor, P.} \& \bibinfo{author}{Jonker, L.}
\newblock \bibinfo{title}{Evolutionary stable strategies and game dynamics}.
\newblock \emph{\bibinfo{journal}{Math. Biosci.}}
  \textbf{\bibinfo{volume}{40}}, \bibinfo{pages}{145 -- 156}
  (\bibinfo{year}{1978}).

\end{thebibliography}
\end{document}